\documentclass[preprint2]{aastex}
\usepackage{spr-astr-addons}
\usepackage{amsmath}

\RequirePackage{color}

\def\etal{{\em et al.} }

\begin{document}
\title{Accretion Flow Dynamics During 1999 Outburst  of XTE J1859+226 - Modeling of Broadband Spectra 
and Constraining the Source Mass}
\slugcomment{Not to appear in Nonlearned J., 45.}
\shorttitle{Nature of Accretion Flows}
\shortauthors{Nandi et al.}

\author{A. Nandi\altaffilmark{1}} \author{S. Mandal\altaffilmark{2}} \author{H. Sreehari\altaffilmark{1,3}} 
\author{Radhika D.\altaffilmark{4}} \author{S. Das\altaffilmark{5}} \author{I. Chattopadhyay\altaffilmark{6}} 
\author{N. Iyer\altaffilmark{7}} \author{V. K. Agrawal\altaffilmark{1}} \and \author{R. Aktar\altaffilmark{5}}
\altaffiltext{1}{Space Astronomy Group, ISITE Campus, ISRO Satellite Center, Outer Ring Road,
Marathahalli, Bangalore, 560037, India}
\altaffiltext{2}{Indian Institute of Space Science and Technology, Trivandrum, 695547, India}
\altaffiltext{3}{Indian Institute of Science, Bangalore, 560012, India}
\altaffiltext{4}{Department of Physics, Dayananda Sagar University, Bangalore, 560093, India}
\altaffiltext{5}{Indian Institute of Technology Guwahati, Guwahati, 781039, India}
\altaffiltext{6}{ARIES, Manora Peak, Naintal, 263002, India}
\altaffiltext{7}{Albanova University Centre, KTH PAP, Stockholm, 10691, Sweden}

\begin{abstract}

We examine the dynamical behavior of accretion flow around XTE J1859+226 during the 1999 outburst 
by analyzing the entire outburst data ($\sim$ 166 days) from RXTE Satellite. Towards this, we 
study the hysteresis behavior in the hardness intensity diagram (HID) based on the 
broadband ($3 - 150$ keV) spectral modeling, spectral signature of jet ejection and the 
evolution of Quasi-periodic Oscillation (QPO) frequencies using the two-component advective 
flow model around a black hole. We compute the flow parameters, namely Keplerian accretion 
rate (${\dot m}_d$), sub-Keplerian accretion rate  (${\dot m}_h$), shock location ($r_s$) and 
black hole mass ($M_{bh}$) from the spectral modeling and study their evolution along 
the q-diagram. Subsequently, the kinetic jet power is computed as 
$L^{\rm obs}_{\rm jet}\sim 3 - 6 \times 10^{37}$ erg~s$^{-1}$ during one of the observed radio 
flares which indicates that jet power corresponds to $8-16\%$ mass outflow rate from 
the disc. This estimate of mass outflow rate is in close agreement with the change in 
total accretion rate ($\sim 14\%$) required for spectral modeling before and during the 
flare. Finally, we provide a mass estimate of the source XTE J1859+226 based on the spectral 
modeling that lies in the range of $5.2 - 7.9 M_{\odot}$ with 90\% confidence.
\end{abstract}


\section{Introduction}

In the quest of the accretion-ejection phenomena in the realm of strong gravity,
astrophysical black holes systems are considered to be the ideal laboratories. Moreover,
most of the galactic black hole 
sources (GBHs) exhibit rich `spectro-temporal' variabilities in X-rays that carry the
signature of accretion-ejection activity \citep{Mirabel-Rodriguez94,
Feroci-etal99,nan01,Belloni2002,Fender-etal09,ADN2015,Aktar-etal2017,Sree18JOAA}.
Extensive X-ray observations by Rossi-X-ray Timing Explorer (RXTE) revealed a direct correlation 
between the evolution of Quasi-Periodic Oscillation (QPO) frequencies and the spectral energy 
distribution in black hole X-ray binaries (XRBs). In case of outbursting sources, it is observed 
that the QPO frequency increases during the rising phase as the source 
transits from hard to intermediate states and subsequently QPOs disappear during soft 
states \citep[and references therein]{Belloni2002, Belloni2005,cha08,nan12,rad14,Iyer-etal15a}.
On the other hand, a reverse trend of QPO frequency variation is observed during the declining
phase of the outburst before the source transits to its quiescent phase 
\citep[references therein]{Belloni2005,nan12,Debnath-etal2013}. In general,  
the hardness intensity diagram (HID) of GBH sources during the outburst exhibits a unique 
feature in the form of a `q-shape' hysteresis loop known as `q-diagram' 
\citep{mc03, HB2005,rem06,nan12,rad14}. In general, GBHs also display weak radio activities 
\citep{F01,CF02} in low-hard states (LHS) and hard-intermediate states (HIMS) during the 
rising phase of the outburst. In addition, a correlation is also observed between the radio 
and X-ray luminosities for several black hole sources \citep{Corbel03,FG2014}. 
Relativistic ejections in the form of radio flares are observed when the source 
transits from its hard-intermediate state (HIMS) to the soft-intermediate state 
(SIMS) \citep{Brock2002,Corbel03,Corbel04,FBG04,HB2005,Fender-etal09,MJ2012}. 
These relativistic ejections ($i.e.,$ jets) are associated either with type A/B 
QPOs \citep{FBG04,Fender-etal09,Sol08} or no QPO with a signature of soft spectrum 
\citep{SV2001,rad14,rnvs16}. Interestingly, type C QPOs are generally observed in both LHS and 
HIMS \citep{Casella2004,HB2005} during the outbursting phase of the
black hole sources.

Numerous theoretical and phenomenological attempts were made to understand the physical 
mechanisms responsible for the above mentioned X-ray characteristics. The evolution of QPO 
frequencies has been interpreted mostly either by means of the propagation of fluctuations 
associated with the hot inner flow \citep{ID2009,ID2011} or due to the oscillation of the post 
shock flow \citep[and references therein]{cha08,Iyer-etal15a}. In addition, the nature of 
the spectral properties of the accretion disc around a black hole are modeled using a 
standard Keplerian disc \citep{sha73} coupled with the `Compton corona'
\citep{SZ94,DK06}. Usually, the `Compton corona' comprises of a hot and dense electron cloud that 
reprocesses the soft photons (i.e., thermal emission) emitted from the Keplerian disc via 
inverse-Comptonization process to produce the high energy component of the radiation spectrum. 
All these studies were carried out considering a static `Compton corona' \citep{TanakaLewin1995}.
A self-consistent attempt has been made using a dynamical two component advective flow 
model \citep{cha95, cha06} while explaining the evolution of the spectral as well as the temporal 
behaviour of the outbursting GBH sources \citep{Iyer-etal15a, Debnath-etal16}.
Recently, \citet{pout17} reported a truncated disc geometry where a disc and hot plasma 
coexists. But, the two component advective flow model deals with a more general scenario where the 
accretion disc is having two components, a sub-Keplerian halo along with a standard Keplerian 
disc and the central hot corona can be produced from the sub-Keplerian halo 
component \citep{GiriChakrabarti-13}.

In order to examine the hysteresis behavior of the outbursting sources, \citet{Mey09}
studied the q-diagram considering the advection dominated accretion flow (ADAF) model, 
whereas \citet{mc10} investigated the same using two-component advective flow model. All these 
studies were performed without taking care of self-consistent modeling of both spectral and 
temporal features related to q-diagram. In addition, the disc-jet coupling associated with 
the q-diagram, mostly observed in soft intermediate state (SIMS), were also ignored. Meanwhile, 
in the context of the disc-jet connection, efforts were made to estimate the jet kinetic power 
for several GBHs considering the phenomenological models \citep{Fender-etal09}. Eventually, 
there appears contradictory claims on the role of black hole spin in powering the 
jets \citep[and references therein] {fgr10,Stei13}, although the launching of jets seems to 
be viable irrespective to the black hole spin parameter \citep{ADN2015,ck16}.
Overall, a complete investigation considering the evolution of the broadband energy spectra, QPO 
frequencies and spectral signature of disc-jet connection for the outbursting
black hole sources still remains a task to be undertaken.

Motivating with this, in the present work, we model the broadband energy ($3 - 150$ keV)
spectra (span over $\sim$ 166 days) of XTE J1859+226 during 1999 outburst considering 
two-component advective flow model around a black hole. Here, we examine the overall hysteresis 
behaviour (in HID) of the source considering the accretion-ejection mechanism. While doing this, 
we theoretically compute the kinetic jet power corresponding to a radio flare 
(observed as jet on MJD 51479.94) and find that the calculated mass loss is in agreement with 
the change in accretion rate required to model the broadband spectral data.  
Moreover, we model the evolution of QPO frequencies while estimating the
size of the post-shock region which is consistent with the best fit spectral
parameters (namely, shock location, see \S 3.3). Finally, based on our recently developed
broadband spectral model \citep{Iyer-etal15a}, we constrain the mass of the black hole
source under consideration. 

In the next section, we describe our model along with the details of observation and analysis 
of the source XTE J1859+226 during 1999 outburst. In \S 3, we present the results 
and estimate the mass of the source. In \S 4, we discuss our results and finally present 
conclusion. 

\section{Model and Observation}\label{s:obs}

\subsection{Model Description}\label{s:mdes}

We consider a geometrically thin accretion disc around a Schwarzschild
black hole. The disc is assumed to be comprised of two-component advective
flows where high angular momentum viscous Keplerian flow \citep{sha73} resides at the
disc equatorial plane and is flanked by the low angular momentum weakly viscous sub-Keplerian 
flow \citep{cha95,wu02,smi07}. In order to satisfy the inner
boundary conditions imposed by the black hole horizon, the accreting flow must be
transonic in nature \citep{AbramowiczZurek81}. Hence, the adopted `hybrid' accretion 
flow must coalesce together to become sub-Keplerian before entering onto the black
hole \citep{GiriChakrabarti-13}. The radial velocity of a Keplerian flow always remain
subsonic, whereas depending on the input parameters, a sub-Keplerian accretion flow may contain 
multiple sonic points \citep{Chakrabarti-89}. Overall, the subsonic flow at the outer
edge of the disc starts accreting towards the black hole due to gravity and becomes supersonic 
after crossing the outer sonic point (usually forms far away from the black 
hole; \citet{Das-etal01}). Rotating flow continues its journey further towards the horizon 
although a virtual barrier around the black hole is developed due to the centrifugal
repulsion that finally triggers a discontinuous transition of the flow in the form of 
a shock wave \citep{Fukue-87,Chakrabarti-89}. At the shock, pre-shock supersonic flow jumps to 
subsonic branch where the flow kinetic energy is converted into thermal energy. Hence, due to shock, 
post-shock flow becomes hot and dense compared to the pre-shock flow and it behaves 
like a `Compton cloud', which we refer as post-shock corona (hereafter, PSC). Finally, the 
subsonic post-shock flow becomes supersonic after passing through the inner sonic point and
eventually enters into black hole. We assume that the Keplerian flow terminates at the shock 
and the PSC is fully sub-Keplerian in nature. The soft photons from the optically thick 
Keplerian disc are intercepted by the optically thin PSC and inverse-Comptonized to produce 
hard X-ray power-law distribution \citep[and reference therein]{cha95, cha06}. We calculate 
the radiation spectrum from both Keplerian and sub-Keplerian components of
the accretion flow to model the observed broadband radiation spectra \citep{Iyer-etal15a}.  
Interestingly, PSC may oscillate due to resonance oscillation \citep{mol96} and/or for a critical 
viscosity that perturbs the shocked flow \citep{lee11,das14} and exhibits QPOs. Finally, following
the prescription of \citet{cha00}, in this work, the evolution of QPO frequencies
in GBH sources is modeled based on the propagatory oscillating shock solution 
(POS) \citep{cha08,Chakrabarti-etal09,nan12}.

In order to describe the space-time geometry around a Schwarzchild black hole, here we adopt a 
pseudo-Newtonian potential given as $\Phi(r)=-1/(r-1)$ \citep{Paczynski-Wiita80}. This pseudo 
potential allows us to solve the governing flow equations following the Newtonian approach 
while retaining all the salient features of strong gravity. Here, $r$ denotes
the radial distance measured in units of Schwarzchild radius, $r_g = 2GM_{bh}/c^2$
where $G$, $M_{bh}$ and $c$ are the gravitational constant, black hole mass and
speed of light, respectively.

\subsection{Salient features of XTE J1859$+$226}

We select the Galactic black hole source XTE J1859 + 226, which has undergone outburst 
phase only during 1999 in the entire RXTE era. 
The coordinated RXTE campaign of the source enables us to study the HID profile
which exhibits spectral state transitions and displays the presence of different types of 
QPOs \citep[and references therein]{Casella2004,HB2005,rad14}. During the same outburst, the 
source has been observed with multiple radio jet ejections which are relativistic in 
nature \citep{Brock2002}.

An orbital period of $6.58\pm0.05$ h and radial velocity amplitude of $541\pm70$ km sec$^{-1}$ 
have been obtained based on the optical photometry and spectroscopy of the binary system. 
Based on this, the mass function of the companion is obtained as $4.5\pm0.6 M_{\odot}$, and for 
an assumed inclination angle $70^{\circ}$, the lower limit of the dynamical mass of this source 
has been estimated as $5.42M_{\odot}$ \citep{Corral-santana-etal01}. The previous estimate of 
the source mass was reported to be $7.7\pm1.2$M$_{\odot}$ \citep{shapo09}. Hence, it is clear 
that there are uncertainties in the estimate of the source mass. This encourages us to study the 
spectral, temporal and ejection characteristics of the source XTE J1859 + 226 and subsequently 
we constrain the mass of the source. 

\subsection{Observation and Analysis}

We analyze the public archival data of RXTE satellite for the 1999 outburst of XTE J1859+226. 
Pointed observations since October 9, 1999 to March 23, 2000 that span over $\sim 166$ days have 
been considered for the present work.

Spectral and temporal analysis are carried out using PCA and HEXTE data in $3 - 150$ keV energy 
band. We follow the standard data reduction procedures as discussed in detail 
in \citet{Casella2004,nan12,rnvs16}. For temporal analysis, we compute the power spectra of all 
data sets in the rising phase of the outburst. A combination of Lorentzian features are used 
for the modeling of the power spectra in order to find out the centroid frequencies 
of QPOs \citep{Casella2004} which display a monotonic increase in frequency during the rising 
phase of the outburst. These QPOs are of C-type in nature and are used to model the evolution of QPOs which 
is presented in \S 3.5. We model the broadband energy spectra (PCA: $3 - 20$ keV and 
HEXTE: $20 - 150$ keV) using phenomenological models consisting of {\it diskbb} and 
{\it powerlaw} components. The fluorescent Fe line at 6.4 keV is modeled using a {\it Gaussian} 
of width $0.7$ keV. From the phenomenological spectral modeling, we estimate the `unabsorbed' 
flux (in units of $10^{-9}$ erg cm$^{-2}$ s$^{-1}$) in $3 - 6$ keV, $6 - 20$ keV and $3 - 20$ keV 
energy bands for generating the HID as shown in Fig. 4a (marked as blue stars). Finally,
we model the broadband energy spectra based on two-component flow model (see \S 3.1) and 
compute the flux ($3-20$ keV) and hardness ratio (HR: ratio of flux in $6-20$ keV to $3-6$ keV) 
to reproduce the model HID profile.

\begin{figure}[h]
\begin{center}
\includegraphics[width=0.4\textwidth,angle=270]{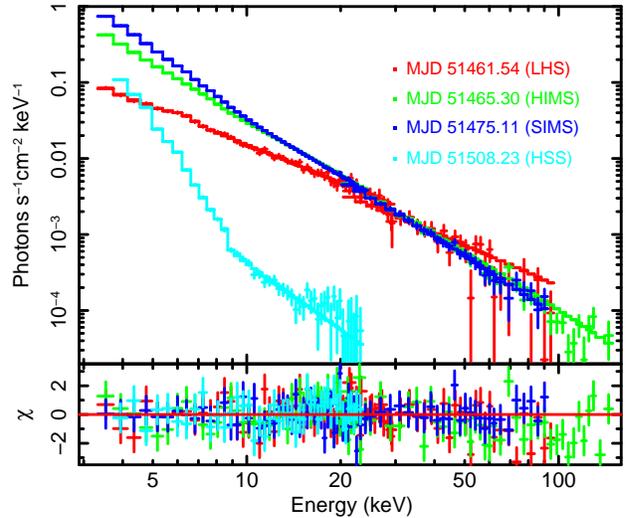}
\end{center}
\caption{Broadband energy spectra ($3 - 150$ keV) using PCA and HEXTE data of RXTE 
fitted by two-component flow model for different states (LHS, HIMS, SIMS and HSS). Residuals are 
shown in bottom panel. See text for details.}
\end{figure}

\section{Modeling and Results}\label{s:res}

We model the broadband energy spectra in the energy band of $3 - 150$ keV using two-component 
advective flow model scheme implemented in {\it XSPEC} \citep{Iyer-etal15a}. 
We estimate the modeled flux in $3 - 6$ keV, $6 - 20$ keV and $3 - 20$ keV bands to reproduce 
the `q-diagram' for the 1999 outburst of XTE J1859+226. We also study the spectral signature of 
jets, using broadband spectral modeling before and after the detection of the radio flare. We 
calculate the jet kinetic power and mass outflow rate for an advective sub-Keplerian disc
\citep{ADN2015,nmdc15}. In addition, we model the evolution of QPO frequencies in 
the rising phase of the outburst (i.e., in LHS \& HIMS), which generally are of C-type QPOs. 

\begin{table*}
        \centering
        \caption{Best fit parameters required to model the broadband energy spectra ($3-150$ keV) 
in various spectral states}
        \begin{tabular}{|c|c|c|c|c|c|c|} 
                \hline
              Day & State & Shock location & Sub-Keplerian  & Keplerian  & Mass & $\chi ^2$/dof \\
                      &  & ($r_s$) & rate, $\dot{m}_h ({\dot M}_{Edd})$ & rate, $ 
\dot{m}_d ({\dot M}_{Edd})$  & ($M_{\odot}$)    &\\\hline
                  1  &       LHS      & $151.0 \pm 28.2$   & $0.310 \pm 0.003$ & $0.125 \pm 0.008$  &  $5.79 \pm 0.24$  &  $98.85/76$\\  
                5.3  &HIMS& $36.5 \pm  3.0$  & $0.135 \pm 0.001$ & $1.630 \pm 0.150 $ &$5.70 \pm 0.13$ &117.57/79\\
                15.4 &SIMS& $20.0 \pm  0.4$ & $0.142 \pm 0.002$ & $1.900 \pm 0.058 $ &$6.00 \pm 0.11$ &46.48/65\\
               48   &HSS    & $5.5 \pm 0.5$    & $0.048 \pm 0.001$ & $0.390 \pm 0.030$ &  $5.90 \pm 0.45$     & 32.85/39\\
                \hline
     \end{tabular}
\end{table*}

\subsection{Modeling of Broadband Energy Spectra} \label{ss:spec}

In two-component advective flow model \citep{cha95,cha06}, we solve the hydrodynamic equations 
in presence of dissipative processes (radiative cooling) for the sub-Keplerian 
component of the flow and treat the Keplerian disc as the source of soft blackbody photons only.
We calculate the energy density of intercepted soft blackbody photons at PSC from every annuli of 
the Keplerian disc. The intercepted photon energy density is used to calculate the cooling 
efficiency due to inverse-Comptonization in the PSC. Moreover, for a set of flow input 
parameters, we obtain the flow variables, namely the number density, electron temperature and 
proton temperature, by solving the governing hydrodynamic equations \citep{MandalChakrabarti2005}. 
Subsequently, these flow variables are employed to calculate the accretion disc radiation spectrum.

We have implemented the above model in {\it XSPEC} \citep{Iyer-etal15a} with four input parameters, 
namely black hole mass ($M_{bh}$, in units of $M_{\odot}$), shock location 
($r_s$, in units of $r_g$), Keplerian ($\dot m_d$) and sub-Keplerian ($\dot m_h$) mass accretion 
rates (in units of Eddington rate). Here, the soft photons (controlled by $\dot m_d$) are supplied 
by the Keplerian disc and the high energy part of the spectrum is generated due to the 
inverse-Comptonization of these soft photons by hot electrons in the PSC which is determined
by $r_s$ and $\dot m_h$. In this model, the high energy and low energy part of the radiation 
spectrum are not independent rather determined by the flow hydrodynamics and coupled through 
the flow input parameters. 

During the 1999 outburst of XTE J1859+226, the source evolved from hard to soft state via 
intermediate states, and finally decays to the hard state \citep{Casella2004,rad14}. We fit 
the broadband spectra ($3 - 150$ keV) for the entire outburst (whenever data is available). 
In Fig. 1, we show the broadband model fitting of: hard state (LHS by red curve) observed 
on \texttt{MJD 51461.54}, hard intermediate (HIMS by green curve) on \texttt{MJD 51465.30}, soft
intermediate (SIMS by blue curve) on \texttt{MJD 51475.11} and soft state (HSS by cyan curve) on
\texttt{MJD 51508.23}. We use {\it phabs} to model the inter-stellar extinction.
Moreover, we also use smeared edge ({\it smedge}) and {\it Gaussian} to model
intrinsic absorption and iron line signatures respectively to obtain the best fit of spectral parameters,
whenever required. In the fitting, we keep a fixed hydrogen column density, $n_H = 0.2\times10^{22}$ $atoms~cm^{-2}$ 
and Gaussian line energy $\sim$ 6.4 keV \citep[and references therein]{rad14}. The residuals of 
the model fitted spectra are shown in the bottom panel of Fig. 1 with $\chi^2$/dof as 
$98.85/76$ (LHS), $117.57/79$ (HIMS), $46.48/65$ (SIMS) and $32.85/39$ (HSS), respectively. 
Best fit parameters required to model the broadband energy spectra in different spectral 
states (days) are presented in Table 1.

The overall spectrum is the combination of the soft and hard components and the relative 
normalization is determined by the fraction of soft photons intercepted by the PSC.
The overall normalization is dependent on the source 
inclination angle and distance and it must be a constant for a given source. The expression for 
the norm value is proportional to $\cos \theta/D^2$ where D is in units of 10 kpc and $\theta$ 
is the angle of inclination of the system. We fix the normalization corresponding to the source
to estimate the mass of the source using two-component flow model fitting and adopt the
following methodology to obtain the normalization value for the source. 
First, we fit the source spectrum from different states considering the norm as a free parameter. 
In reality, a physical system is more complex than an ideal one and hence normalization varies 
among the data sets. But, the obtained normalization values from various data sets do not 
vary dramatically. Hence, our best approximation of the norm value is the average of these 
individual norm values. Finally, we refit all the data with the average norm value to get the 
individual mass estimations and combine them to obtain the better mass estimate.

The broadband modeling shows (Fig. 1) that low energy photon flux increases and spectral 
state becomes steeper as the source transits into the soft state. The low energy flux in HSS is
relatively low as the source has entered to the decay phase of the outburst (see Fig. 5).
Finally, in high soft state (HSS), the high energy contribution becomes very weak as the soft 
photons from the Keplerian disc are able to completely cool the hot electrons in PSC.

\subsection{Confidence Intervals to Estimate the Source Mass} \label{ss:q-dia}

As noted in the previous section, we have multiple observations of the source in different 
spectral states and the spectrum of each observation is being modeled 
independently. The mass 
of the central object is a fit parameter which is estimated from the spectral fit.
Thus, we have multiple estimates of the mass of the central object, which can 
be treated as independent measurements to be combined to place a better constraint on the mass 
of the source. Recently, \citet{Iyer-etal15a} estimated the mass of black hole from broadband 
spectral modeling of the X-ray spectrum. We follow the same procedure in order to combine
the individual mass measurement and estimate a confidence interval. 

\begin{figure}[h]
  \centering
  \includegraphics[width=0.48\textwidth]{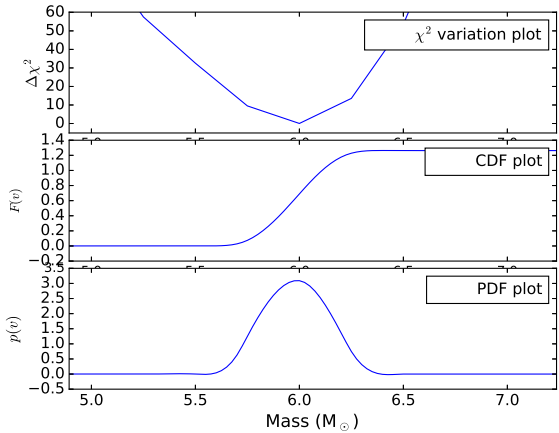}
  \caption{Variation of $\Delta \chi^2$, CDF and PDF as function of mass obtained 
from the spectral fit of SIMS state. 
  The $\chi^2$ variations about the parameter minima are used to obtain 
confidence intervals from which the CDF and PDF are derived. See text for details.}
  \label{fig:prob}
\end{figure}

\begin{figure}[h]
\begin{center}
\includegraphics[width=0.475\textwidth,angle=0]{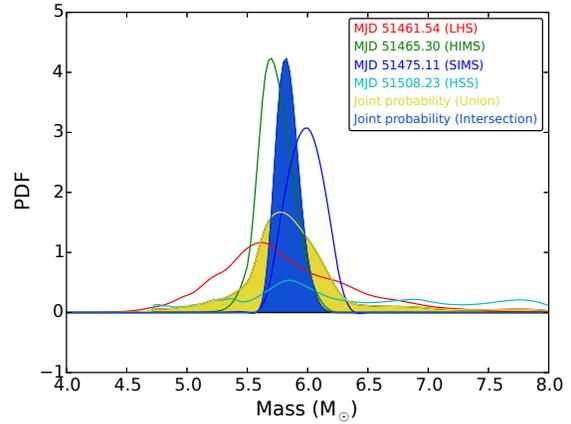}
\end{center}
\caption{PDFs obtained from individual observations of different spectral states (same with Fig. 1) 
as marked in the plot. Probable mass range for final Union and Intersection PDFs are shown as 
shaded regions. See text for details.}
\end{figure}

The basic premise for combining mass estimates from individual estimates comes from the 
application of naive Bayes theorem. In order to do this, the probability density function (PDF) 
of the mass value from each observation is required. To find the PDF of the mass value, we use 
the variation in the fit parameter ($\chi^2$) about the minima of the fit along with 
variation in the mass value. This can be obtained using {\it steppar} command of XSPEC and 
it gives the confidence value ($C_{v_1}^{v_2}$) that the mass lies in a given 
interval ($v_1$, $v_2$) for a given change in $\chi^2$ ($\Delta_1$). By evaluating these 
confidence intervals for different values of $\Delta$, we can numerically obtain the 
cumulative distribution function (CDF = $F(v)$) of the mass \citep{Iyer-etal15a,Iyer-etal15b}. 
From the CDF, we can then obtain the PDF ($p(v)$) of the mass. These steps are listed in
equations 1-\ref{eqn:prob} and illustrated in Fig. 2 for the SIMS spectrum of XTE J1859 + 226.

\begin{eqnarray}
  C_{v_1}^{v_2} = \int_{v_1}^{v_2} p(v)dv = P(\chi^2 > \Delta_1), \\
  A_{v_1}^{v_{min}} = \frac{v_{min} - v_1}{v_2 - v_1}C_{v_1}^{v_2}; \; 
A_{v_{min}}^{v_2} = \frac{v_2 - v_{min}}{v_2 - v_1}C_{v_1}^{v_2}, \\
  F(v) = \int_{-\infty}^{v} p(v)dv = \begin{cases}
	A_{-\infty}^{v_{min}} - A_{v}^{v_{min}}, & v < v_{min} \\
	A_{-\infty}^{v_{min}} + A_{v_{min}}^v, & v > v_{min}.
  \end{cases}
  \label{eqn:prob}
\end{eqnarray}

Combining these independent PDFs (obtained from fitting of different spectral states data, 
see Fig. 1), we then estimate an overall bound on the value of mass. This is done by using a 
naive Bayes approach of multiplying the individual PDFs to obtain the joint PDF which is the 
intersection of individual PDFs. We also obtain an estimate of the worst case upper and lower 
bound on the mass value by calculating the Union of PDFs as is enumerated 
in \citet{Iyer-etal15b}. The final mass bounds is then stated as the 90\% confidence limits on 
each of these approaches, with the intersection based on PDF giving a tighter constrained mass 
value while the union PDF giving the worst case bounds. The intersection of PDFs from 
different data sets put a tight bound on mass but it may not particularly favour any of the PDFs. 
For example, if two PDFs are well separated, still their intersection may provide a narrow PDF 
but this estimate may not be particularly meaningful. On the other hand, if we add the PDFs, 
it provides an overall minimum bound, but the estimate can be poor due to systematics.
This approach gives a worst case mass estimate of the source XTE J1859+226 as 
$5.2 - 7.9M_{\odot}$ with 90\% confidence as shown in Fig. 3 (yellow shaded regions) 
along-with the individual PDFs as obtained from various spectral states (marked in the plot). 
A more tighter mass constraint comes in a narrow range of $5.7 - 6.0M_{\odot}$. The tighter 
bounds can be considered true if there are no systematic errors in the evaluation of mass 
from any one of the spectral states. However, with many unknown systematics being present 
both in the model, and in the data and uncertainties in the response
obtained from RXTE, we stick to quote the worst case bounds as our final limit 
as $5.2 - 7.9M_{\odot}$.

A lower value of mass bounds can be obtained if additional physical processes 
like effects of atomic lines, intrinsic disc absorptions/reflections and jet contributions are included in the model
self-consistently. With this, such a model would be able to consistently fit all observations having single norm. 
Also, to avoid the systematics in the data, one would not be able to combine the results from all 
the states. Rather we may need to consider only LHS and HSS where the systematics due to atomic 
line emissions and jet ejections are comparatively less. 
Even in LHS there can be contribution due to steady winds if not jet.
Other possibility of improvement is that a simultaneous multi-wavelength observations are required to model 
the jet/wind contributions and accordingly this contribution can be removed from the accretion disc.

Further, we attempt to get a mass bound by choosing spectra from 
LHS and HSS where the systematics in the data are low and modeling them simultaneously with tied mass and 
norm. Fig. 4 shows the simultaneous fitting of LHS and HSS spectra of XTE J1859+226 
during its 1999 outburst.
The fit results give a mass of $5.91 \pm 0.28 ~M_{\odot}$ with $\chi^2$/dof
equal to $150.82/104$. This corresponds to an overall uncertainty
of 9.46 \%. This estimate is within the broad range given by the union of PDFs and
is overlapping with the tighter bounds given by the intersection of PDFs. As the intersection 
of PDFs may not favor the maxima of individual PDFs, we consider the estimate based on the
simultaneous fitting with tied norm and mass values from LHS and HSS.

\begin{figure}[h]
\includegraphics[scale=0.32, angle=-90]{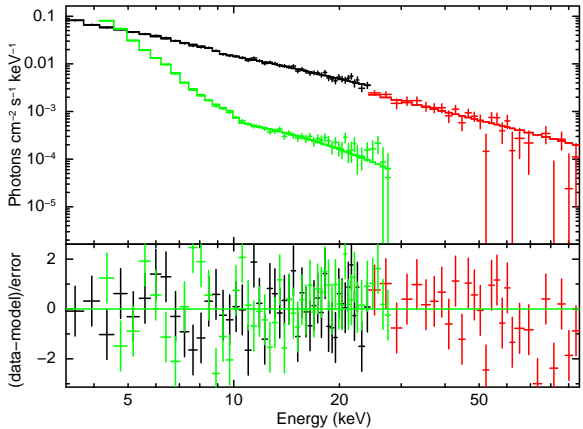}
\caption{Simultaneous fitting of LHS (black-red for PCA and HEXTE data) and HSS (green 
for PCA data) spectra from the 1999 outburst of XTE J1859+226 done by tying the mass 
and norm parameters. Fitted mass is around $5.91 \pm 0.28 ~M_{\odot}$, which is within the broad
range of mass estimated from union of PDFs (see Fig. 3).}
\label{fig:tied} 
\end{figure}

\subsection{Modeling of HID (q-diagram)} \label{ss:mo-q-dia}

In order to model the HID, we fit the entire outburst (span over $\sim$ 166 days) spectral data 
using two-component advective flow model and compute the fluxes in the energy band of 
$3 - 20$ keV, $3 - 6$ keV and $6 - 20$ keV. Upon estimating the fluxes and hardness
ratios, we plot the variation of total flux as function of hardness ratio as shown in Fig. 5a 
(marked with red circles). It is to be noted that the `modeled' profile (red filled circles) is 
in close agreement with the observed (blue stars) HID obtained from the phenomenological 
modeling of the data. We also estimate the model 
parameters (${\dot m}_{h}$, ${\dot m}_{d}$ and $r_s$) from the spectral fitting during the 
entire $1999$ outburst and in Fig. 5b, we present the variation of $\dot m_h$ 
(green filled asterisks), $\dot m_d$ (red filled diamonds) and $r_s$ (blue filled circles), 
respectively. We also plot the total flux observed (light gray) in the same figure. 
It is evident that total flux increases with the total accretion rate and follows the same trend
of accretion rate during the entire outburst.
We observe that as the source moves from hard to soft state in the rising phase, ${\dot m}_{d}$ 
(red filled diamonds) increases and $r_s$ (blue filled circles) moves towards the black hole. 
While in the decay phase, the Keplerian accretion rate (${\dot m}_{d}$) decreases and 
shock ($r_s$) moves outward.
During the outburst, we see an anti-correlation between the two accretion rates which is expected.
Except the first few days and last few days of the outburst, $r_s$ varies between few 
to $30~r_g$ and sub-Keplerian rate ($\dot m_h$) remains $\sim$ 0.1 and hence the outburst dynamics 
is mostly controlled by Keplerian accretion rate $\dot m_d$. Usually, the dynamics of the
Keplerian disc is governed by viscous timescale which is typically around few 
days \citep{mc10} for a disc around a stellar mass black hole and this time scale is consistent 
with the rising phase ($\sim$ 7-8 days) of the outburst of XTE J1859+226. After the rising phase, 
the source continues to move towards the soft state where the supply of Keplerian matter becomes 
significant. Finally, as the source enters into the decaying phase of the outburst, supply of 
sub-Keplerian matter is increased reducing the Keplerian matter supply.

\begin{figure*}
  \includegraphics[width=.475\linewidth, height=7.8cm]{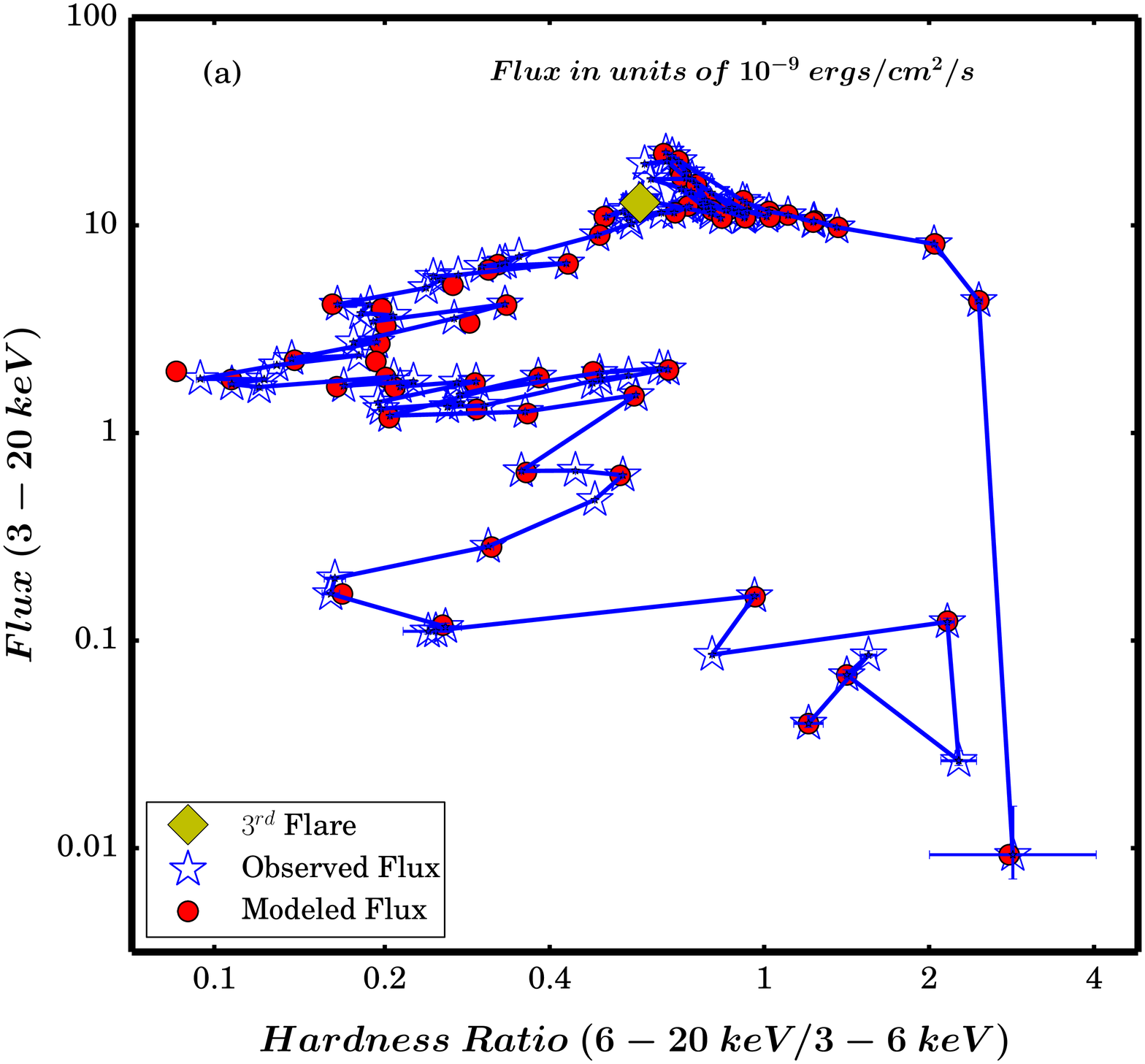}
  \includegraphics[width=.475\linewidth, height=7.8cm]{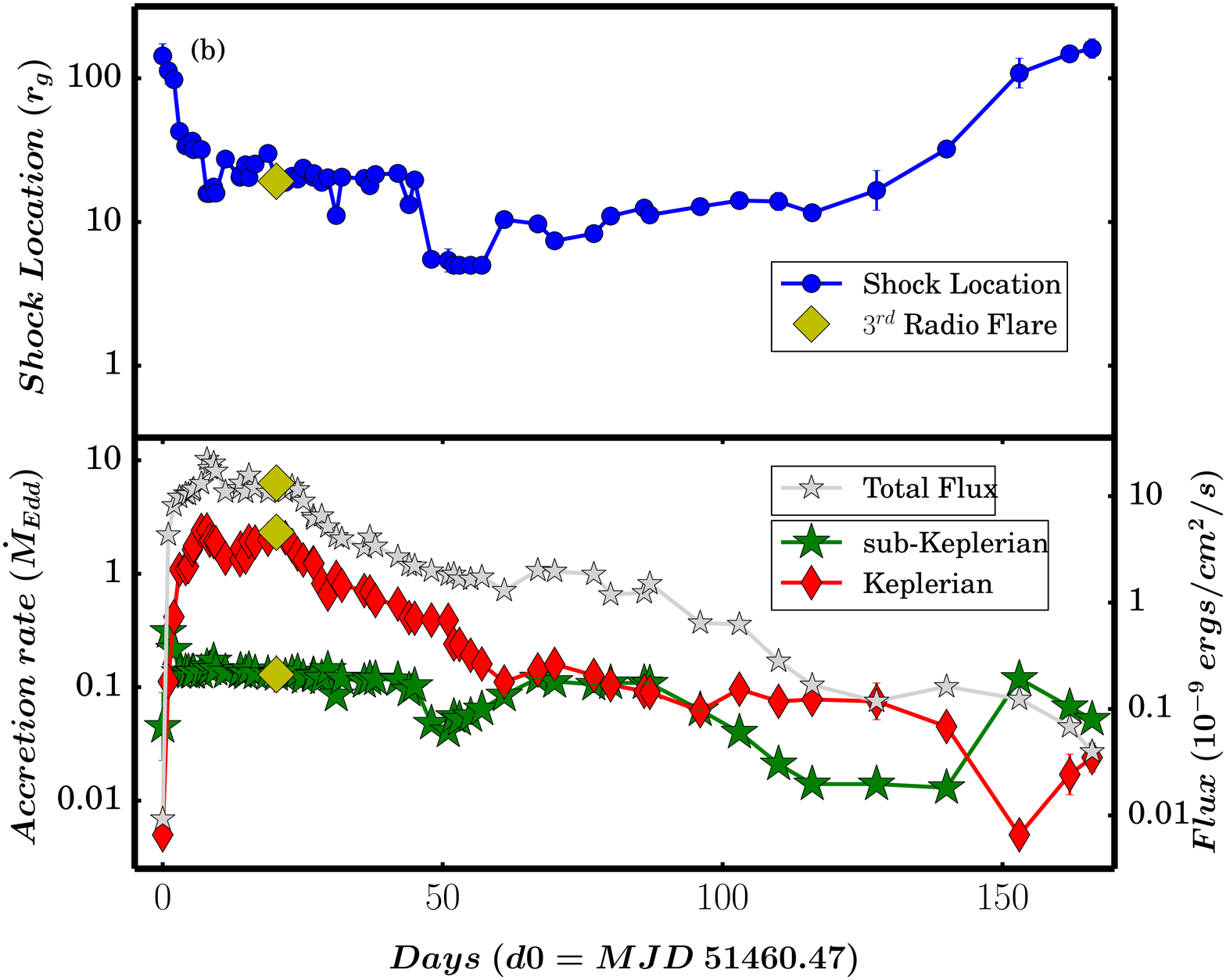}
\caption{(a) HID of the 1999 outburst of XTE J1859+226 plotted using X-ray flux from 
phenomenological model (marked with blue star along-with error bars) and two-component model 
based HID (marked as red filled-circle) are presented. A good agreement is seen between the 
modeled q-profile and the observed q-profile. (b) The time evolution of shock location ($r_s$, 
blue circles in upper panel), sub-Keplerian accretion rate (${\dot m}_h$, green asterisks), 
Keplerian accretion rate (${\dot m}_d$, red diamonds), and total observed flux (light gray stars)
along the q-track are shown (in bottom panel). See text for details.}
\label{}
\end{figure*}

\subsection{Spectral Signature of Disc-Jet Connection and Estimation of Jet Power} \label{ss:disc-jet}

Multiple transient jets ($\equiv$ radio flares) are observed in the $1999$ outburst of 
XTE J1859+226 \citep{Brock2002} although QPOs are not seen in X-ray data \citep{rad14} during the 
radio flares. A general understanding towards this is that multiple transient jets are observed 
in the intermediate state of the outbursting sources and these ejections may be triggered 
during the state transition from HIMS to SIMS. In this work, as a representative case, we choose 
the third radio flare (read as $F_3$) when the source is clearly in the SIMS state.
We use the X-ray spectral data during $F_3$ flare in order to investigate the spectral signature 
of the jet ejections. We model three broadband spectra before (MJD 51478.78 in red), after 
(MJD 51483.11 in blue) and during (MJD 51480.04 in green) the radio flare
\citep{Brock2002,rad14} as depicted in Fig. 6 ($F_3$ is marked in Fig. 5a and 5b as well). 
The spectrum (green curve) during the flare indicates the presence of excess soft photon flux 
whereas the hard X-ray flux is noticeably reduced compared with the other two spectra 
(red and blue curves). During the flare, a two-component broadband spectral modeling of 
the observed data requires an inward movement of shock (smaller PSC) and the increase of 
mass accretion rate ($\sim 14\%$) in comparison with the same needed to model the 
spectral data before the flare. In Table 2, we have summarize the model fitted spectral 
parameters - both accretion rates, shock location and $\%$ change in total rates during the 
F3 radio flare. It is evident that the halo accretion rate remains almost constant across 
the flare and the significant change in mass accretion rate is due to the variation in 
Keplerian disc rate.

\begin{table*}
	\centering
	\caption{Model fitted spectral parameters around the F3 radio flare.}
	\label{tab:flare3}
	\small
	\begin{tabular}{lccccr} 
		\hline
		MJD & Shock location ($r_s$) & Halo rate ($\dot{M}_{Edd}$) & Disc rate ($\dot{M}_{Edd}$) & Total rate ($\dot{M}_{Edd}$)& Change (\%)\\
		\hline
		51478.77 & $30.0 \pm 0.9$ &$0.131 \pm 0.003$ &$2.01 \pm 0.072$& $2.14 \pm 0.072$&-\\
        51480.04 (flare) & $19.2 \pm 0.6$ &$0.129 \pm 0.001$ &$2.32 \pm 0.098$& $2.44 \pm 0.098$&14\\
        51483.11 & $24.8 \pm 0.8$ &$0.132 \pm 0.002$ &$2.20 \pm 0.060$& $2.33 \pm 0.060$&09\\
		\hline
     \end{tabular}
\end{table*}

\begin{figure}[h]
\begin{center}
\includegraphics[width=0.35\textwidth,angle=270]{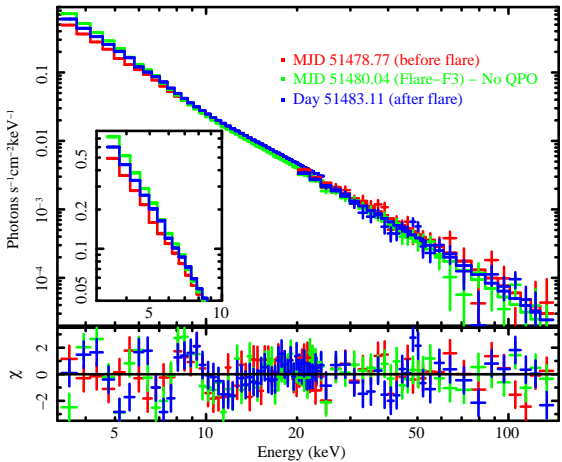}
\end{center}
\caption{Spectral signature of mass-loss during third flare ($F_3$) of the 1999 outburst of 
XTE J1859+226. QPO frequency is not observed during the flare. Maximum change in accretion rate 
is observed to be $\sim 14\%$ between before and during the radio flare. See text for details.
}
\end{figure}

In order to quantify the mass flux of the outflowing matter, we calculate the observed kinetic 
jet power ($L^{\rm obs}_{\rm Jet}$) employing the expression given by \citep{Longair-11},
$$
L_{\rm jet}^{\rm kin} \sim 3\times10^{33}\eta_{\rm jet}^{4/7}\left(\frac{\Delta t}{\rm sec}
\right)^{2/7}\left(\frac{\nu}{\rm GHz}\right)^{2/7}
$$
$$
\times \left(\frac{S_{\nu}}{\rm mJy}\right)^{4/7}
\left(\frac{D}{\rm kpc}\right)^{8/7}~~{\rm erg~sec}^{-1},
\eqno(4)
$$
where $\eta_{\rm jet}$ is the jet efficiency factor and its value is of the order of
unity as estimated from observation \citep{F01}. Here, $\Delta t$ is the rise
time of the jet which we consider as $24$ hours as a representative value. The observed
radio flux is denoted by $S_\nu$, where $\nu$ is the frequency of radio observation and
$D$ is the distance to the source. During the $\rm F_3$ flare, the radio flux is observed
as $S_{\nu} \sim 25~{\rm mJy}$ at $\nu = 3.9~{\rm GHz}$ \citep{Brock2002}. 
When jet moves with speed comparable to the speed of light ($c$),
emitted radiations from the jet are expected to be 
relativistically Doppler boosted towards the observer. In order to account the relativistic
Doppler effect, we introduce a correction factor $f(\Gamma, \theta)$ while estimating the
observed kinetic jet power \citep{F01} as,
$$
L_{\rm jet}^{\rm obs} =f(\Gamma , \theta) \times L_{\rm jet}^{\rm kin}~~{\rm erg~sec}^{-1},
\eqno(5)
$$
where $\Gamma$ is the bulk Lorentz factor and $\theta$ is the inclination angle between the
line of sight and the direction of the jet axis. In this study, we consider the
inclination angle as $\theta \sim 60^{o}-70^{o}$ \citep{Corral-santana-etal01} and the bulk
Lorentz factor as $\Gamma \sim 2-3$ (as jets are assumed to be mildly relativistic),
respectively. 
Employing these values, we estimate the mean value of $f(\Gamma, \theta) = 9.74$ (Fender 2001). 
Using the above data along with a range of distance to the source as
$D = 6-11 ~ {\rm kpc}$ \citep{hynes02,zurita02}, we estimate the observed kinetic jet power to be 
$L^{\rm obs}_{\rm jet}\sim 3-6 \times 10^{37}$ erg~s$^{-1}.$

In the context of accretion-ejection mechanism, rotating accretion flow experiences
virtual barrier ($i.e.$, PSC) in the vicinity of the black hole where a part of the accreting 
matter is deflected in the vertical direction to form bipolar jets
\citep[and references therein]{c99,Chattopadhyay-Das-07,Das-Chattopadhyay-08,ck16}.
Towards this, \citet{ADN2015} recently estimated the mass outflow rate ($R_{\dot{m}}$)
around the rotating black hole in terms of the inflow parameters. Moreover, using the 
RXTE observations of several black hole sources, they calculated the X-ray luminosity
of a given source as $L_x = 4 \pi D^2 F_x$, where $F_x$ is the total unabsorbed X-ray flux.
Following this approach and considering the radiative efficiency of the accreting 
matter $\eta \sim 0.06$ (for non-rotating black holes), in this work, we calculate the accretion 
rate of the black hole as $\dot{m}_{\rm acc} = L_x/\eta c^2$. Subsequently, the model 
estimated jet kinetic power is estimated as \citep{ADN2015},
$$
L_{\rm jet}^{\rm est} = R_{\dot{m}} \times \dot{m}_{acc} \times c^2~~{\rm erg~sec}^{-1},
\eqno(6)
$$
where $R_{\dot{m}}$ represents the mass outflow rate defined as the ratio of outflow and inflow 
mass flux. In this work, using the best fit spectral parameters during flare $F_3$ 
with $r_s \sim 20~r_g$ (Fig. 5b) and the difference
of X-ray fluxes during (MJD 51480.04) and before (MJD 51478.77) the flare 
as $F_x \sim 2.16 \times 10^{-9}$ erg~cm$^{-2}$~s$^{-1}$, we estimate the model kinetic jet 
power $L_{\rm jet}^{\rm est}$ employing equation (6). 
We find that $L_{\rm jet}^{\rm est}$ is in agreement with $L^{\rm obs}_{\rm jet}$ 
provided the mass outflow rate ($R_{\dot{m}}$) lies in the range $\sim 8-16\%$ of the 
accretion rate. This mass outflow rate is consistent with the change in accretion 
rate ($\sim 14\%$) required to fit the broadband spectra (Fig. 6) before and during the 
jet ejection as mentioned earlier.

\subsection{Evolution of QPO Frequency ($\nu$)} \label{ss:evo-qpo}

In this section, we examine the evolution of QPO frequencies during the rising phase of 
the outburst following the Propagating Oscillatory Shock (POS) model (\citealt[]{cha08}, 
\citealt[]{Chakrabarti-etal09}, \citealt[]{nan12}, \citealt[]{Iyer-etal15a}), where shock
location is treated as free parameter. 

The frequency of the QPO which is the rate at which the shock front oscillates is given 
by (\citealt[]{Iyer-etal15a} and references therein),
$ \nu_{qpo}= \dfrac{c}{2\pi R r_g r_s\sqrt{r_s -1}}$, where $R$ is the compression ratio defined 
as the ratio of post-shock to the pre-shock matter densities. In this model, the mass of the 
black hole that appears in the expression of $r_g$ is also a free parameter. Due to the limited 
number of observed data points (see Fig. 7) as well as the limited knowledge of the methodology 
itself, namely how $r_s$ is evolving with time, here we use the black hole mass estimated from 
the broadband spectral modeling and consider it as a fixed parameter. In the rising phase of 
the outburst, the QPO frequency increases with time and we assume that the shock is moving 
inward with an acceleration ($a_t$). Hence, the location of the shock is changing with 
time (t) as $r_s= r_{s0} - (v_ot + 0.5 a_tt^2)/r_g$, where $r_{s0}$ is the initial shock
location and $v_0$ is the initial shock drift speed.
These parameters are estimated from the model fitting of the evolution of QPO frequencies.
The evolution of the QPO frequencies (black points) and its model fitting (magenta curve)
are depicted in Fig. 7. The estimated shock locations are presented by red circles. The bottom 
panel shows the residual of the fitted QPO frequencies. 

In Fig. 7, we over plot the shock locations (blue stars) estimated from spectral 
modeling, which are in close agreement (except the 1st observation) with POS model fitted 
shock locations. We also estimate the QPO frequencies using POS formula considering the model
fitted shock locations. The calculated QPO frequencies are shown in Fig. 7 (green stars).  
The results are summarized in Table \ref{tab:xs_nu}, where comparison between results obtained 
from temporal (POS) and spectral (two-component) modeling are given. 
Here, the best fit requires 
$v_0=20.9$ m\,s$^{-1}$, $r_{s0}=263~ r_g$ respectively. According to our analysis, 
initially the oscillating shock front accelerates inward with
$a_t\sim -1.08$ m\,s$^{-1}$\,day$^{-1}$ up to the shaded region
(both observations show similar spectro-temporal features) which is 
followed by an acceleration $a_t\sim -5.08$ m\,s$^{-1}$\,day$^{-1}$ in the HIMS.
The estimated shock locations ($i.e.$, size of PSC) decreases as the QPO
frequencies (C-types) evolve from LHS to HIMS which is consistent with the 
variation of $r_s$ estimated from the broadband spectral modeling (Fig. 5b; Table 3). 
Since A-type and B-type QPOs observed in SIMS do not show noticeable evolution 
(\citealt{Casella2004,rad14}), we exclude them while modeling the evolution of QPO frequencies.

\begin{figure}[h]
\begin{center}
\includegraphics[width=0.48\textwidth,angle=0]{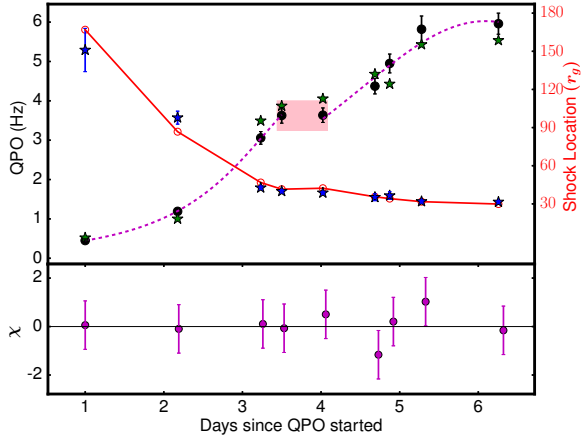}
\end{center}
\caption{Evolution of QPO frequencies (black points) in LHS and HIMS with time (days) of the 
1999 outburst of XTE J1859+226 is fitted with POS model solution (dotted magenta curve). 
The model fitted shock locations (red curve) decreases with the increase of QPO frequencies 
as predicted by the model. Residuals of POS fitting are shown in bottom panel. The 
shock location (blue stars) estimated from two-component flow modeling of energy spectra and 
the corresponding estimated QPO frequencies (green stars) are also shown.
}
\label{fig:QPOevol}
\end{figure}

\begin{table*}
	\centering
	\caption{Comparison of QPO frequencies and shock location estimated from temporal
and spectral modeling. We get $r_{sm}$ and $\nu$ by modeling energy spectra with two-component 
flow model and POS formula, while $r_{so}$ is obtained from POS fitting of observed QPOs.}
	\label{tab:xs_nu}
	\begin{tabular}{l|c|c|c|c|r} 
		\hline
		MJD & QPO (Hz) & FWHM (Hz) & $r_{sm}$ ($r_g$) & $r_{so}$ ($r_g$) & $\nu$(Hz)\\
		\hline
		51461.57 & 0.45 & 0.089 & 151.0 $\pm$ 17.0 & 166.96 & 0.52\\
		51462.76 & 1.19 & 0.187 & 97.80 $\pm$ 5.1& 86.76 & 1.00\\
		51463.83 & 3.05 & 0.378 & 42.70 $\pm$ 1.3& 46.80 & 3.49\\
		51464.10 & 3.62 & 0.462 & 39.90 $\pm$ 1.1& 41.52 & 3.87\\
		51464.63 & 3.63 & 0.433 & 38.70 $\pm$ 0.8& 42.30 & 4.05\\
		51465.30 & 4.37 & 0.464 & 35.20 $\pm$ 2.7& 35.58 & 4.67\\
		51465.49 & 4.94 & 0.562 & 36.50 $\pm$ 3.0& 34.14 & 4.42\\
		51465.90 & 5.81 & 0.790 & 31.90 $\pm$ 1.1& 31.73 & 5.43\\
		51466.89 & 5.96 & 0.630 & 31.50 $\pm$ 0.9& 29.86 & 5.53\\
		\hline
     \end{tabular}
\end{table*}

\begin{figure}[h]
\begin{center}
\includegraphics[width=8.2cm]{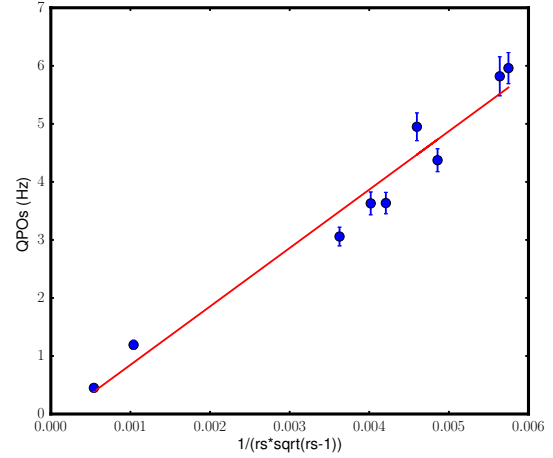} 
\end{center}
\caption{Correlation between QPO frequencies and the POS model function 
(POS formula; \S 3.5) obtained from energy spectral modeling.}
\label{fig:QPO_rep}
\end{figure}

Next, we demonstrate that the two values of frequencies, measured independently from 
timing and spectral modeling, agree quite well. For this, we have plotted observed QPOs against 
$1/(r_s\sqrt{(r_s-1)})$, variable term in POS equation, where $r_s$ is obtained from spectral modeling. 
We have fitted this the variation with a straight line 
(Fig. 8). The slope of the fitting is 1006.43 with an interception of -0.15. Comparing with 
the POS formula, the value of $R*r_g$ is obtained as $47441.16$, which provides a compression ratio 
(R) as 2.87. The estimated R lies in the theoretically accepted limit ($1-4$) for non-relativistic shocks.

\section{Discussion and Conclusions}\label{s:dis}

In this paper, we examine the accretion flow dynamics of the outbursting XTE J1859+226 
source by modeling the spectral and temporal characteristics along its q-track.
Towards this, we theoretically model the broadband ($3-150$ keV) spectral evolution of the 
1999 outburst (span over $\sim$ 166 days) of this source using two-component advective flow 
model. In this analysis, four input parameters, namely Keplerian accretion rate (${\dot m}_d$), 
sub-Keplerian accretion rate  (${\dot m}_h$), shock location ($r_s$) and black hole 
mass ($M_{bh}$), are used to model all the observations.
Apart from this, the normalization parameter of the model is a constant scaling factor 
for a given source. 
We have fitted spectra from various states of the source considering 
norm as a free parameter and then take an average of all these independent norm values. This average norm 
value is used as a constant norm for all the observations of the source under consideration.
The model fitted norm value provides an estimate of the product $\cos \theta/D^2$ and 
does not allow an independent estimate of distance and inclination angle of the source. 
If the inclination angle is assumed to lie between 
$60^0 - 70^0$ \citep{Corral-santana-etal01} for this
source, the distance is calculated as $4.8 - 5.8$ kpc which is in agreement with the 
distance range $4.6-8$ kpc reported by \citet{hynes02}.

Following our approach as mentioned in \S 3.1 and \S 3.3, we reproduce the q-profile of 
the source which is in agreement with the observed q-diagram (Fig. 5a).
Moreover, our results suggest that during the rising phase of the outburst, Keplerian accretion rate 
increases and shock front moves inwards (Fig. 5b) while the sub-Keplerian accretion rate 
initially decreases and becomes almost constant afterwards. On the contrary, towards the end of 
the outburst both the accretion rates tend to readjust where sub-Keplerian counter part exceeds 
the Keplerian component.

In the decay phase, a reverse trend in the variation of accretion rates is seen and PSC reconciles
the hysteresis property of this source. Also, the broadband spectral modeling indicates that as
the source moves towards softer states, the low energy X-ray flux increases with a steepening of
high energy power law (Fig. 1). 
Further, we constrain the mass of the source using broadband spectral modeling by generating
PDFs (Fig. 3) obtained from different spectral states as discussed in \S 3.2. The mass
range is estimated as $5.2 - 7.9~M_{\odot}$ with 90\% confidence 
which is above the dynamically measured lower mass limit \citep{Corral-santana-etal01}.
We quote this larger mass range as the effect of systematics in model and data. 
Model systematics include effects 
from atomic lines, winds and jet emissions and these need to be self-consistently
included in the model. To check the systematics in data, we fit LHS and HSS data considering 
tied norm and mass  simultaneously and following this, we obtain the mass
estimate as $5.91 \pm 0.28 ~M_{\odot}$ with an overall uncertainty
of 9.46 \%.

During the radio flare, the reduction of hard X-ray flux is observed (green curve in Fig. 6),
as a part of the accreting matter is ejected from the PSC in the form of jets (see \S 3.4). Hence,
the size of PSC is reduced that allows the Keplerian disc to move inwards enabling the
increase of Keplerian accretion rate. With this, the availability of soft photons is enhanced
significantly. A few hours after the flare (blue curve in Fig. 6), shock recedes outward
and the contribution of hard X-ray flux increases. 
A similar characteristics is also noticed during all the radio flares observed
in this source \citep{rad14}, as well as other black hole sources \citep{SV2001,nan01,rnvs16}. 
In addition, we compute the kinetic jet power ($L_{\rm jet}^{\rm obs}$) employing 
the observed radio flux during the third flare and  $L_{\rm jet}^{\rm obs}$ is then
used to estimate the mass outflow rate ($R_{\dot m}$) which is found to be
consistent with the change in accretion rate (${\dot m}_h + {\dot m}_d$) required
to fit the spectral data before and during the radio flare.

Employing the POS solution (\S 3.5), it is ascertained that the shock
decelerates inward in the initial phase and further slows down to
a smaller value in SIMS phase. This evidently indicates that QPO frequency 
increases during the rising phase of the outburst (Fig. 7). Moreover, we
find that shock location calculated from the POS model is consistent with 
the size of the corona ($i.e.$, PSC) as obtained  from spectral modeling (Fig. 5b).  
In Fig 8. we show the correlation of QPO frequencies with the functional form 
($r_s$ estimated from spectral modelling) of the POS equation. 
The slope of the correlation (after fitting a straight line) provides a compression ratio (R)
of around 2.87, well within the expected range.

Spectro-temporal variability similar to the source XTE J1859 + 226 is also observed 
in multiple outbursts of the source GX 339-4 \citep[and references therein]{nan12} 
and in other outbursting GBH sources \citep{rem06}. The investigation of these 
variabilities observed in other GBH sources and constraining the mass of the sources
are under progress and will be reported elsewhere.

\acknowledgments

Authors are thankful to the reviewer for his/her valuable suggestions and 
comments that helped to improve the quality of the manuscript.
This research has made use of data obtained through the High Energy Astrophysics Science Archive 
Research Center (HEASARC) online service, provided by the NASA/Goddard Space Flight Center.
AN thanks GD, SAG; DD, PDMSA and Director, ISAC for encouragement and continuous support 
to carry out this research.



\begin{thebibliography}{}

\bibitem[Abramowicz \& Zurek (1981)]{AbramowiczZurek81}
Abramowicz, M. A. \& Zurek, W. H., 1981, \apj, 246, 314

\bibitem[Aktar \etal (2015)]{ADN2015}
Aktar, R., Das, S., \& Nandi, A., 2015, \mnras, 453, 3414

\bibitem[Aktar \etal (2017)]{Aktar-etal2017} 
Aktar, R., Das, S., Nandi, A., \& Sreehari, H.\ 2017, \mnras, 471, 4806 

\bibitem[Belloni \etal (2002)]{Belloni2002}
Belloni, T., Psaltis, D., \& van der Klis, M., 2002, \apj, 572, 392 

\bibitem[Belloni \etal (2005)]{Belloni2005}
Belloni, T., Homan, J., Casella, P., et al., 2005, \aap, 440, 207

\bibitem[Brocksopp \etal (2002)]{Brock2002}
Brocksopp, C., et al., 2002, MNRAS, 331, 765

\bibitem[Casella \etal (2004)]{Casella2004}
Casella, P., Belloni, T., Homan, J., \& Stella, L., 2004, A\&A, 426, 587

\bibitem[Chakrabarti \& Titarchuk (1995)]{cha95} 
Chakrabarti, S K., \& Titarchuk, L., 1995, \apj, 455, 623. 

\bibitem[Chakrabarti (1999)]{c99}
Chakrabarti, S. K.,  1999, \aap, 351, 185

\bibitem[\protect\citeauthoryear{Chakrabarti \& Manickam}{2000}]{cha00}
Chakrabarti, S. K., \& Manickam, S. G., 2000, \apj, 531, L41

\bibitem[Chakrabarti \& Mandal (2006)]{cha06}
Chakrabarti, S. K., \& Mandal, S., 2006, \apjl, 642, L49

\bibitem[Chakrabarti \etal (2008)]{cha08}
Chakrabarti, S. K., Debnath, D., Nandi, A., \& Pal, P. S., 2008, \aap, 489, L41

\bibitem[Chakrabarti, Dutta, \& Pal (2009)]{Chakrabarti-etal09}
Chakrabarti, S. K., Dutta, B. G., \& Pal, P. S., 2009, MNRAS, 394, 1463 

\bibitem[Chakrabarti (1989)]{Chakrabarti-89} 
Chakrabarti, S. K., 1989, ApJ, 347, 365 

\bibitem[Chattopadhyay \& Das (2007)]{Chattopadhyay-Das-07}
Chattopadhyay, I., \& Das, S., 2007, \na, 12, 454

\bibitem[Chattopadhyay \& Kumar(2016)]{ck16} 
Chattopadhyay, I., \& Kumar, R.\ 2016, \mnras, 459, 3792

\bibitem[Corbel \& Fender (2002)]{CF02}
Corbel, S., \& Fender, R. P., 2002, \apj, 573, L35

\bibitem[Corbel \etal (2003)]{Corbel03}
Corbel, S., Nowak, M., Fender, R. P., et al., 2003, \aap, 400, 1007

\bibitem[Corbel \etal (2004)]{Corbel04}
Corbel, S., Fender, R. P., Tomsick, J. A., et al., 2004, \apj, 617, 1272

\bibitem[Corral-Santana \etal (2011)]{Corral-santana-etal01}
Corral-Santana, J. M., et al., 2011, \mnras, 413, 15

\bibitem[Das \etal (2001)]{Das-etal01}
Das, S., Chattopadhyay, I., \& Chakrabarti, S. K., 2001, ApJ, 557, 983 

\bibitem[Das \& Chattopadhyay (2008)]{Das-Chattopadhyay-08}
Das, S., \& Chattopadhyay, I., 2008, \na, 549, 556

\bibitem[Das \etal (2014)]{das14}
Das, S., Chattopadhyay, I., Nandi, A., \& Molteni, D., 2014, \mnras, 442, 251

\bibitem[Debnath et al.(2013)]{Debnath-etal2013} 
Debnath, D., Chakrabarti, S.~K., \& Nandi, A.\ 2013, Advances in Space Research, 52, 2143

\bibitem[Debnath \etal (2016)]{Debnath-etal16}
Debnath, D., Chakrabarti, S.K., \& Mondal, S., 2016, \mnras, 440, 121

\bibitem[Done \& Kubota (2006)]{DK06}
Done, C., \& Kubota, A., 2006, \mnras, 371, 1216

\bibitem[Feroci \etal (1999)]{Feroci-etal99}
Feroci, M., Matt, G., Pooley, G., et al.\ 1999, \aap, 351, 985

\bibitem[Fender (2001)]{F01}
Fender, R. P., 2001, \mnras,  322, 31 

\bibitem[Fender \etal (2004)]{FBG04}
Fender, R. P., Belloni, T., \& Gallo, E., 2004, \mnras, 355, 1105

\bibitem[Fender \etal (2009)]{Fender-etal09}
Fender, R. P., Homan, J., \& Belloni, T., 2009, \mnras, 396, 1307

\bibitem[Fender \etal (2010)]{fgr10}
Fender, R. P., Gallo, E., \& Russell, D., 2010, \mnras, 406, 1425

\bibitem[Fender \& Gallo (2014)]{FG2014}
Fender, R., \& Gallo, E., 2014, \ssr, 183, 323 

\bibitem[Fukue (1987)]{Fukue-87}
Fukue, J., 1987, PASJ, 39, 309

\bibitem[Giri \& Chakrabarti (2013)]{GiriChakrabarti-13}
Giri, K., \& Chakrabarti, S. K., 2013, MNRAS, 430, 2836

\bibitem[Homan \& Belloni (2005)]{HB2005}
Homan, J., \& Belloni, T., 2005, \apss, 300, 107 

\bibitem[Hynes \etal (2002)]{hynes02}
Hynes, R. I., Haswell, C. A., Chaty, S., Shrader, C. R. \& Cui, W., 2002, \mnras, 331, 169-179

\bibitem[Iyer \etal (2015a)]{Iyer-etal15a}
Iyer, N., Nandi, A., \& Mandal, S., 2015a, \apj, 807, 108 

\bibitem[Iyer \etal (2015b)]{Iyer-etal15b}
Iyer, N., Nandi, A., \& Mandal, S., 2015b, 
Astronomical Society of India Conference Series, 12, 97

\bibitem[Ingram \etal (2009)]{ID2009}
Ingram, A., Done, C., \& Fragile, P. C., 2009, \mnras, 397, L101

\bibitem[Ingram \& Done (2011)]{ID2011}
Ingram, A., \& Done, C., 2011, \mnras, 415, 2323 

\bibitem[Longair (2011)]{Longair-11}
Longair, S. M., 2011, High Energy Astrophysics (Cambridge University Press)

\bibitem[Lee \etal (2011)]{lee11}
Lee, S. J., Ryu, D., \& Chattopadhyay, I., 2011, \apj, 728, 142

\bibitem[Maccarone \& Coppi (2003)]{mc03}
Maccaroni, T. J., \& Coppi, P.S., 2003, \mnras, 338,189

\bibitem[Mandal \& Chakrabarti(2005)]{MandalChakrabarti2005}
Mandal, S., \& Chakrabarti, S.~K.\ 2005, \aap, 434, 839

\bibitem[Mandal \& Chakrabarti (2010)]{mc10}
Mandal, S., \& Chakrabarti, S. K., 2010, \apjl, 710, L147

\bibitem[Meyer-Hofmeister \etal (2009)]{Mey09}
Meyer-Hofmeister, E., Liu, B. F., \& Meyer, F., 2009, \aap, 508, 329 

\bibitem[Miller-Jones \etal (2012)]{MJ2012}
Miller-Jones, J. C. A., Sivakoff, G. R., Altamirano, D., et al., 2012, \mnras, 421, 468 

\bibitem[Mirabel \& Rodr{\'{\i}}guez(1994)]{Mirabel-Rodriguez94}
Mirabel, I.~F., \& Rodr{\'{\i}}guez, L.~F.\ 1994, \nat, 371, 46

\bibitem[Molteni \etal (1996)]{mol96}
Molteni, D., Sponholz, H., \& Chakrabarti, S. K., 1996, \apj, 457, 805

\bibitem[Nandi \etal (2001)]{nan01}
Nandi, A., Chakrabarti, S. K., Vadawale, S. V., \& Rao, A. R., 2001, \aap, 380, 245

\bibitem[Nandi \etal (2012)]{nan12}
Nandi, A., Debnath, D., Mandal, S., \& Chakrabarti, S. K., 2012, \aap, 542, A56

\bibitem[Nandi \etal (2015)]{nmdc15}
Nandi, A., Mandal, S., Das, S., \& Chattopadhyay, I., 2015, Astronomical Society of India Conference Series, 12, 69

\bibitem[Paczy\'nski \& Wiita (1980)]{Paczynski-Wiita80}
Paczy\'nski, B., \& Wiita, P. J., 1980, \aap, 88, 23.

\bibitem[Poutanen \etal (2017)]{pout17}
Poutanen, J., Veledina, A., Zdziarski, A. A., 2017, arXiv 1711.08509

\bibitem[Radhika \& Nandi (2014)]{rad14}
Radhika, D., \& Nandi, A., 2014, AdSpR, 54, 1678

\bibitem[Radhika \etal (2016)]{rnvs16}
Radhika, D., Nandi, A., Agrawal, V. K., \& Seetha, S., 2016, MNRAS, 460, 4403

\bibitem[Remillard \& McClintock (2006)]{rem06}
Remillard, R. A., \& McClintock, J. E., 2006, \araa, 44, 49

\bibitem[Shakura \& Sunyaev (1973)]{sha73}
Shakura, N. I., \& Sunyaev, R. A., 1973, \aap, 24, 337

\bibitem[Shaposhnikov \& Titarchuk (2009)]{shapo09}
Shaposhnikov, N., \& Titarchuk, L., 2009, \apj, 699, 453

\bibitem[Smith \etal (2007)]{smi07}
Smith, D. M., Dawson, D. M., \& Swank, J. H., 2007, \apj, 669, 1138

\bibitem[Soleri \etal (2008)]{Sol08}
Soleri, P., Belloni, T., \& Casella, P., 2008, \mnras, 383, 10989

\bibitem[Sreehari \etal (2018)]{Sree18JOAA}
Sreehari, H., Nandi, A., Radhika, D., Iyer, N. \& Mandal, S., J.~Astrophysics.~Astronomy, 39, arXiv 1802.05163

\bibitem[Steeghs \etal (2013)]{Stee13}
Steeghs,D., J. E. McClintock, S. G. Parsons, M. J. Reid, S. Littlefair, \& V. S. Dhillon, 2013, \apj, 768, 185

\bibitem[Steiner \etal (2013)]{Stei13}
Steiner, J. F., McClintock, J. E., \& Narayan, R., 2013, \apj, 762, 104 

\bibitem[Svensson \& Zdziarski (1994)]{SZ94}
Svensson, R., \& Zdziarski, A. A., 1994, \apj, 436, 599 

\bibitem[Tanaka \& Lewin(1995)]{TanakaLewin1995}
Tanaka, Y., \& Lewin, W.~H.~G.\ 1995, X-ray Binaries, 126

\bibitem[Vadawale \etal (2001)]{SV2001}
Vadawale, S. V., Rao, A. R., Nandi, A., \& Chakrabarti, S. K., 2001, \aap, 370, L17 

\bibitem[Wu \etal (2002)]{wu02}
Wu, K., et al. 2002, \apj, 565, 1161

\bibitem[Zurita \etal (2002)]{zurita02}
Zurita, C., S{\'a}nchez-Fern{\'a}ndez, C., Casares, J., et al., 2002, \mnras, 334, 999 


\end{thebibliography}
\end{document}